\documentclass{article}
\usepackage{microtype}
\usepackage{booktabs}
\usepackage{xcolor}
\usepackage{hyperref}

\usepackage[accepted]{mlsys2025}

\makeatletter
\AtBeginDocument{\def\mlsys@appearing{Preprint.}}
\makeatother

\usepackage{xspace}
\usepackage{graphicx}
\usepackage{tikz}

\usepackage{algpseudocode}
\usepackage{amsmath,amssymb}
\usepackage{amsthm}
\newtheorem{theorem}{Theorem}[section]
\newtheorem{proposition}[theorem]{Proposition}
\newtheorem{lemma}[theorem]{Lemma}

\usepackage{caption}
\usepackage{subcaption}
\usepackage{float}
\usepackage{enumitem}
\usepackage{listings}
\usepackage{mdframed}
\usepackage{booktabs,threeparttable,pifont}
\usepackage{xcolor}
\usepackage{hyperref}
\usepackage{booktabs,multirow}
\usepackage{booktabs}
\usepackage[table]{xcolor}
\usepackage{tabularx}
\usepackage{capt-of}
\usepackage{xurl}

\newenvironment{denseitemize}{
\begin{itemize}[leftmargin=1.5em]
}{\end{itemize}}

\def\name{OneReel\xspace}

\definecolor{RefRed}{HTML}{B00000}
\definecolor{LinkBlue}{HTML}{0000FF}

\hypersetup{
    colorlinks=true,
    linkcolor=LinkBlue,
    citecolor=RefRed,
    urlcolor=LinkBlue,
    pdfsubject={}
}

\begin{document}

\twocolumn[
\mlsystitle{OneReel: Efficient Multi-Scene Video Generation with Scene-Structured Attention}
\mlsyssetsymbol{equal}{*} 
\begin{mlsysauthorlist}
\mlsysauthor{Wei Zhang}{equal,uiuc}
\mlsysauthor{Yuchen Wang}{equal,uiuc}
\mlsysauthor{Tianrui Feng}{uta}
\mlsysauthor{Hongrui Wu}{stanford}
\mlsysauthor{Chenfeng Xu}{uta}
\mlsysauthor{Fan Lai}{uiuc}
\end{mlsysauthorlist}
\mlsysaffiliation{uiuc}{University of Illinois Urbana-Champaign, Urbana, IL, USA}
\mlsysaffiliation{uta}{University of Texas at Austin, Austin, TX, USA}
\mlsysaffiliation{stanford}{Stanford University, Stanford, CA, USA}
\mlsyscorrespondingauthor{Wei Zhang}{zhangw2@illinois.edu}
\mlsyscorrespondingauthor{Fan Lai}{fanlai@illinois.edu}
\mlsyskeywords{Video generation, inference systems, scheduling}
\vskip 0.3in
\begin{abstract}
Multi-scene video generation must follow scene-specific prompts while preserving consistency across scenes. Dense joint generation enables cross-scene information exchange but incurs quadratic attention cost, whereas separate generation avoids this cost at the expense of cross-scene consistency. We present \name, a training-free system that exploits scene structure to enable efficient multi-scene generation. Our key insight is that multi-scene generation can preserve complete intra-scene attention while restricting cross-scene information exchange to a bounded shared memory. \name uses scene prompts and durations to establish scene-specific attention scopes and dynamically selects a bounded set of representative latent frames for cross-scene sharing. This reduces video attention complexity from quadratic to linear in the number of scenes. To translate these savings into execution speedups, \name packs scene-local and shared K/V into dense variable-length attention and employs work-balanced multi-GPU partitioning with communication-aware selection between K/V gathering and Ulysses-style parallelism. Our evaluations across four video generation models show that \name achieves $1.7$--$4.9\times$ end-to-end speedup over dense attention and $1.2$--$2.2\times$ over existing sparse-attention advances, while improving text-following accuracy by 34--37 percentage points and maintaining cross-scene consistency.
\end{abstract}
]
\printAffiliationsAndNotice{\mlsysEqualContribution}

\section{Introduction}
\label{sec:introduction}

Recent advances in video generation models, such as Wan3.0~\cite{Wan3.0}, CogVideoX~\cite{Cogvideox}, and MiniMax H3~\cite{MinimaxH3}, are enabling increasingly complex long-form media synthesis and filmmaking~\cite{Seedance}. These applications require more than generating longer clips: users need to control individual scenes (shots), while maintaining consistency across them. In multi-scene generation, a request specifies a sequence of scene prompts and durations~\cite{MultiShotMaster}. The number and durations of the scenes define the scene structure. The generated video should follow this structure while achieving both scene-specific text-following accuracy and cross-scene consistency.

\begin{figure*}[t]
  \centering
  \includegraphics[width=\textwidth]{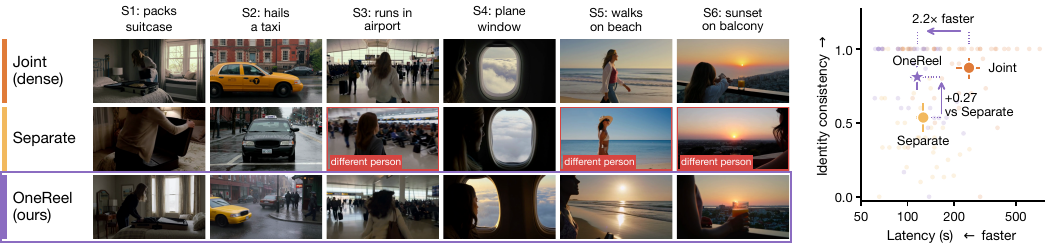}
  \vspace{-.6cm}
  \caption{\name uses scene-structured attention to generate multi-scene videos faster than dense joint generation while achieving higher cross-scene identity consistency than separate generation.}
  \vspace{-.4cm}
  \label{fig:teaser}
\end{figure*}

Existing models can generate multi-scene videos in two ways, but neither simultaneously provides efficient generation and effective cross-scene coordination. \emph{Separate generation} independently generates each scene from its own prompt and concatenates the resulting clips~\cite{Storymem}. Although shared images or conditioning on the preceding scene can help maintain cross-scene consistency~\cite{Wan2.1, StoryDiffusion}, it requires additional generation passes or training. In contrast, \emph{joint generation} denoises all scenes as a single sequence, allowing information to propagate across scenes through global attention~\cite{Holocine}. However, this unrestricted interaction incurs quadratic attention cost in the total video length and does not distinguish scene-specific content from information that should be shared.

Sparse attention can reduce this computational cost, but does not fully exploit the structure of multi-scene generation. Existing training-free methods select attention scopes through online profiling or attention scores~\cite{SparseVideoGen,SparseVideoGen2,AdaSpa}, or use predefined spatial-temporal patterns~\cite{RadialAttn, FastVideo}. To preserve generation quality, many of these methods retain dense attention during the initial denoising steps and enable sparse execution only afterwards, yet early dense steps already account for 20--53\% of total denoising time (\S\ref{subsec:structure-opportunity}). 

Our key insight is that multi-scene generation can separate complete scene-local attention from bounded cross-scene information exchange. Scene prompts and durations provide a structural prior before denoising begins, which imply which video tokens belong to each scene and which text description should condition them. This motivates \emph{scene-structured sparsity}: retain intra-scene attention for scene-specific content, and share only a small amount of information across scenes for consistency. As such, each scene attends to all of its own tokens and to a limited set of tokens from other scenes, thus removing unnecessary cross-scene interactions.

However, turning scene-structured attention scopes into fast, consistent generation raises unique challenges. 
\emph{First, how can cross-scene information sharing remain both effective and scalable?} Even restricting sharing to scene boundaries incurs quadratic attention cost as the number of scenes grows, since every scene accesses information from all others. Bounding the shared context reduces this cost but risks losing information needed for cross-scene consistency. The challenge is to maintain a compact, informative memory as scene count increases and video representations evolve throughout denoising.
\emph{Second, how can structured sparsity translate into efficient GPU execution?} Each scene combines dense local attention with a small set of foreign information (e.g., K/V) scattered across the video sequence. Executing these irregular scopes with sparse kernels can introduce substantial overhead.
\emph{Third, how can multi-GPU execution balance computation without excessive communication?} Scenes have different lengths and attention scopes, so equal-sized query partitions need not have equal computational costs. Distributing every scene across GPUs can balance attention work but requires remote K/V exchange even for intra-scene attention.

We introduce \name, a training-free runtime that translates scene structure into efficient attention computation and multi-GPU execution. It addresses these challenges through three mechanisms.
\textbf{(1) Scene-structured attention with bounded frame memory.} \name assigns each scene complete local video attention and its own text conditioning. For cross-scene information exchange, it clusters pooled latent-frame representations across the video and selects a bounded set of representative frames. Each scene attends to its complete local context and selected foreign frames, preserving local detail while bounding cross-scene attention work (\S\ref{sec:scene-attention}).
\textbf{(2) Dense execution of structured attention.} Rather than directly executing irregular sparse attention masks, \name copies the bounded foreign K/V next to each scene's local K/V. This converts the retained attention interactions into contiguous, variable-length dense scopes. By replicating only the shared context, \name trades limited data movement for substantially higher kernel efficiency (\S\ref{subsec:packing}).
\textbf{(3) Work-balanced multi-GPU execution.} \name partitions queries according to their attention workloads while preserving complete scenes whenever possible, thus reducing inter-GPU communication. A request-level communication cost model then selects between locality-aware K/V gathering and Ulysses-style parallelism~\cite{USP}, accounting for the aggregate communication requirements of the resulting GPU placement (\S\ref{subsec:balance}).

We implement \name atop xDiT~\cite{xDiT} and evaluate it on Wan2.2~\cite{Wan2.1}, HunyuanVideo~\cite{HunyuanVideo}, MiniMax H3~\cite{MinimaxH3} and CogVideoX1.5~\cite{Cogvideox}. Across workloads with varying scene structures, \name achieves end-to-end speedups of $1.7$--$4.9\times$ over dense attention and $1.2$--$2.2\times$ over existing sparse-attention advances. It further improves text-following accuracy by 34–37 percentage points while maintaining cross-scene consistency (\S\ref{sec:eval}).

Overall, this paper makes the following contributions:
\begin{denseitemize}
  \item We identify scene structure as a training-free prior for decomposing multi-scene attention into local computation and bounded cross-scene exchange (\S\ref{sec:motivation}--\S\ref{sec:scene-attention}).

  \item We introduce novel mechanisms to enable scene-structured attention for diverse model architectures, bounded frame memory, efficient dense scope packing, and locality-aware multi-GPU execution (\S\ref{sec:design}). 

  \item We demonstrate \name's substantial end-to-end speedup while improving generation quality (\S\ref{sec:eval}).
\end{denseitemize}

\section{Background and Motivation}
\label{sec:motivation}

\subsection{The Cost--Consistency Trade-off in Multi-Scene Generation}
\label{subsec:cost-quality}
 
A multi-scene request can be served by an existing video diffusion model in two training-free ways (Fig.~\ref{fig:teaser}): (i) \emph{Joint} generation denoises all scenes as one sequence conditioned on the request's scene prompts, allowing scenes to exchange information through attention at every denoising step; and (ii) \emph{Separate} generation denoises each scene with its own prompt and concatenates the results. These two modes differ in both execution cost and their ability to combine scene-specific content with cross-scene consistency.
 
\paragraph{Joint generation spends most of its time on cross-scene attention.}
Let a video contain $L$ video tokens partitioned into $K$ scenes, where scene $i$ contains $n_i=\rho_i L$ tokens and $\sum_i \rho_i=1$. Dense video attention computes $L^2$ query--key interactions, while intra-scene attention covers $\sum_i(\rho_i L)^2$ query--key pairs. The remaining $L^2-\sum_i(\rho_i L)^2$ pairs connect different scenes. With $K$ equally sized scenes, cross-scene pairs account for $1-1/K$ of all video attention pairs, regardless of what the scenes actually share. 

\begin{figure}[t]
    \centering
    \begin{minipage}{0.49\linewidth}
        \centering
        \includegraphics[width=\linewidth]{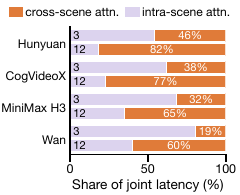}
        \caption{Joint generation spends most of its time on cross-scene attention.}
        \label{fig:cost}
    \end{minipage}
    \hfill
    \begin{minipage}{0.49\linewidth}
        \centering
        \includegraphics[width=\linewidth]{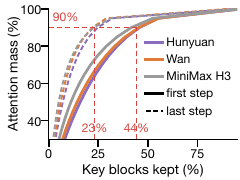}
        \caption{Early attention is too diffuse for a small sparse budget.}
        \label{fig:coverage}
    \end{minipage}
\end{figure}

This computational growth translates into substantial execution overhead. Fig.~\ref{fig:cost} compares joint and separate generation at a fixed duration per scene. Across four models, the excess attention time of joint generation accounts for 19--46\% of its total latency with three scenes, rising to 60--82\% with twelve. Real-world videos can contain tens of scenes~\cite{SF20K},  further increasing the cost of dense cross-scene attention.

\textbf{Separate generation avoids this global cost but loses consistency.}
Removing cross-scene attention lets each scene be generated in isolation, conditioned only on its own prompt. This provides no exchange of video representations across scenes, and recurring characters are re-imagined in every scene. In Fig.~\ref{fig:teaser} (left), the traveler changes identity across separately generated scenes. Across the benchmark prompts (right), the share of scene pairs that DINOv2 judges to show the same subject falls from 0.87 for joint generation to 0.54 for separate generation. Joint generation keeps identities stable, but its scenes often blend into one another and drift from their own prompts, so it is not a quality ceiling either.

These results highlight two distinct requirements. Each scene must preserve its own prompt-specific content, yet scenes that share visual elements need a means of exchanging information. Subject similarity alone does not establish that both requirements have been satisfied. Therefore, the systems challenge is not simply to eliminate cross-scene attention, but to retain useful information exchange without paying the cost of unrestricted global interactions.

\begin{table}[t]
  \centering
  \small
  \begin{threeparttable}
    \caption{Sparse attention spends 20--53\% of denoising time in dense warm-up.}
    \label{tab:sparse-methods}
    \begin{tabular}{@{}lcc@{}}
      \toprule
      Method & Warm-up steps & Warm-up time\tnote{a} \\
        \midrule
    SVG~\cite{SparseVideoGen}    & 5 / 50  & 22.2\% \\
    SVG2~\cite{SparseVideoGen2}  & 5 / 50  & 20.3\% \\
    Sol-Attn~\cite{SolAttn}      & 10 / 50 & 43.0\% \\
    Radial~\cite{RadialAttn}     & 12 / 50 & 44.2\% \\
    STA~\cite{FastVideo}         & 15 / 50 & 52.9\% \\ 
      \bottomrule
    \end{tabular}
    \begin{tablenotes}\footnotesize
      \item[a] Share of denoising time spent in the dense steps.
    \end{tablenotes}
  \end{threeparttable}
\end{table}

\begin{figure}[t]
  \centering
  \begin{minipage}{\linewidth}
    \centering
    \includegraphics[width=\linewidth]{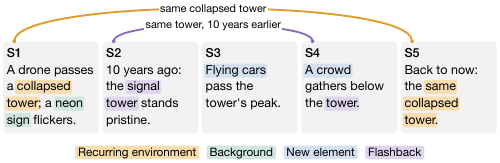}
    \caption{A request specifies scene boundaries and per-scene prompts. Which content recurs is implicit and must be discovered at runtime.}
    \label{fig:scene-structure}
    \vspace{-.3cm}
  \end{minipage}
\end{figure}
 
\subsection{Scene Structure Provides an Attention Prior}
\label{subsec:structure-opportunity}
 
\paragraph{Attention-derived sparsity is difficult to exploit early in denoising.}
Sparse attention reduces the cost of joint generation without removing all cross-scene interactions. Existing training-free methods select interactions using attention scores~\cite{AdaSpa,SparseVideoGen2} or predefined spatial-temporal patterns~\cite{RadialAttn, FastVideo}. However, even with a sparse pattern available, applying it throughout denoising can affect generation quality. A common practice is to retain dense attention during the initial steps and enable sparse execution only afterwards. Using their evaluated configurations, we notice that this dense warm-up occupies the first 10--30\% of denoising steps but accounts for 20--53\% of denoising time (Tab.~\ref{tab:sparse-methods}). Fig.~\ref{fig:coverage} further illustrates this challenge. We partition the attention matrix into fixed-size blocks, each covering a group of query--key interactions. During early denoising, attention mass is broadly distributed across these blocks, so even retaining the highest-scoring blocks requires a large fraction of them. Moreover, this diffuse distribution persists across denoising steps, making aggressive early sparsification prone to discarding substantial attention mass and degrading generation quality. So, early denoising provides little concentration for score-based sparsification to exploit.

\paragraph{Scene structure provides an attention prior before denoising.}
Unlike attention-score-based sparsity, multi-scene requests provide structural information before generation begins. Scene durations determine which video tokens belong to each scene, while scene prompts specify their corresponding text conditioning (Fig.~\ref{fig:scene-structure}). This information identifies intra-scene attention regions without relying on intermediate attention scores. Although the request does not specify which cross-scene video representations are needed for consistency, it enables a natural decomposition: preserve complete attention within each scene while selectively sharing a bounded amount of information across scenes. Importantly, this structure is available from the first denoising step, even when attention scores are too diffuse to support aggressive score-based sparsification.

\section{Scene-Structured Attention}
\label{sec:scene-attention}

We next introduce how \name decomposes multi-scene attention into complete scene-local computation and bounded cross-scene information sharing. We first describe how this decomposition maps to different model architectures using a unified \emph{scene-structured attention} (\S\ref{subsec:scene-attention}), then how the shared memory is constructed and bounded (\S\ref{subsec:sharing}).

\subsection{Scene-Structured Attention across Architectures}
\label{subsec:scene-attention}

Given a request with $K$ scenes, \name aims to associate each scene with its text conditioning while allowing selected video information to be shared across scenes. Let $V_i$ denote the video tokens assigned to scene $i$ by its duration, and $T_i$ the tokens of its prompt. Scene boundaries determine which queries belong to each scene, but do not require those queries to access only local keys. Our assignment distinguishes \emph{query ownership}, determined by scene boundaries, from \emph{attention scope}, which determines which keys each query can access.

For a given attention head, let $\mathcal{S}_i$ denote the set of video keys accessible to video queries in scene $i$. It contains the full set $V_i$ and a bounded subset of foreign video tokens selected by the sharing mechanism in Section~\ref{subsec:sharing}. This abstraction specifies the logical information flow independently of the physical arrangement of tokens in memory.

\begin{figure}[t]
  \centering
  \includegraphics[width=\columnwidth]{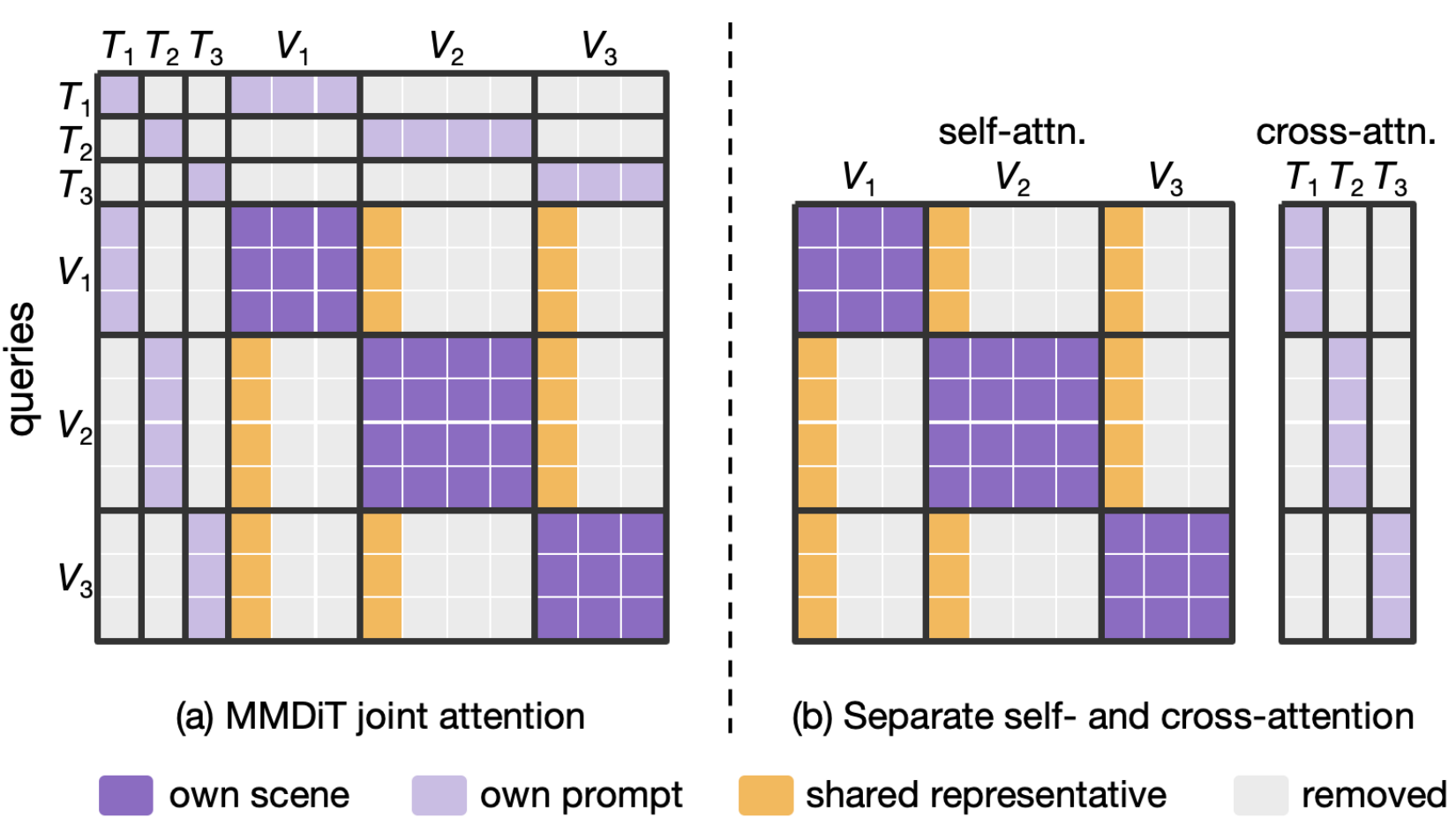}
  \caption{Scene-structured attention across model architectures. Rows denote queries and columns denote keys. Purple denotes scene-local interactions; orange denotes selected foreign video K/V from a shared memory. The example shows one head; text queries remain local.}
  \label{fig:scope-matrix}
\end{figure}

\paragraph{Supporting different model architectures.}
As shown in Fig.~\ref{fig:scope-matrix}a, for MMDiT-style model architectures like HunyuanVideo, text and video tokens participate in joint attention. \name restricts video queries from scene $i$ to its prompt tokens, its own video tokens, and selected video tokens from other scenes. Specifically, for each scene $i$, \name allows video queries to attend to $
  \mathcal{K}^{\mathrm{video}}_i
  = \mathcal{S}_i \cup T_i$,
while its text queries attend only to $
  \mathcal{K}^{\mathrm{text}}_i
  = V_i \cup T_i$.
Thus, video queries may access selected foreign video representations, but text queries remain restricted to their corresponding scene.

In models such as Wan2.2, video self-attention and text--video cross-attention are separate operations (Fig.~\ref{fig:scope-matrix}b). \name applies the video-reading rule in self-attention and restricts cross-attention to the corresponding scene prompt. Therefore, both architectures apply the same logical principle: complete local video context, scene-specific text conditioning, and selective cross-scene video access. A shared video key may be read by multiple scenes, but each query produces exactly one output at its original position.

\subsection{Cross-Scene Information Sharing}
\label{subsec:sharing}
Scene-local attention preserves scene-specific information, but recurring subjects and scene transitions require cross-scene sharing.  Allowing every scene to read tokens from all others restores quadratic attention complexity. Even restricting sharing to a few frames per scene retains quadratic growth in scene count when every scene accesses all shared frames. Instead, \name bounds the total shared context using a memory of representative latent frames. The challenge is to keep this memory compact without discarding information useful for cross-scene consistency.

\paragraph{Frame-level representation and selection.}
Rather than selecting individual tokens, \name selects complete latent frames as the units of cross-scene sharing. Each representative frame retains all its spatial tokens, preserving its complete spatial representation for receiving scenes. To identify representative frames, \name constructs a compact feature for each latent frame. During the conditional pass of denoising step $t$, it mean-pools the video hidden states at a designated selection layer over the spatial tokens within each frame and normalizes the resulting features.

\begin{figure}[t]
  \centering
  \includegraphics[width=\columnwidth]{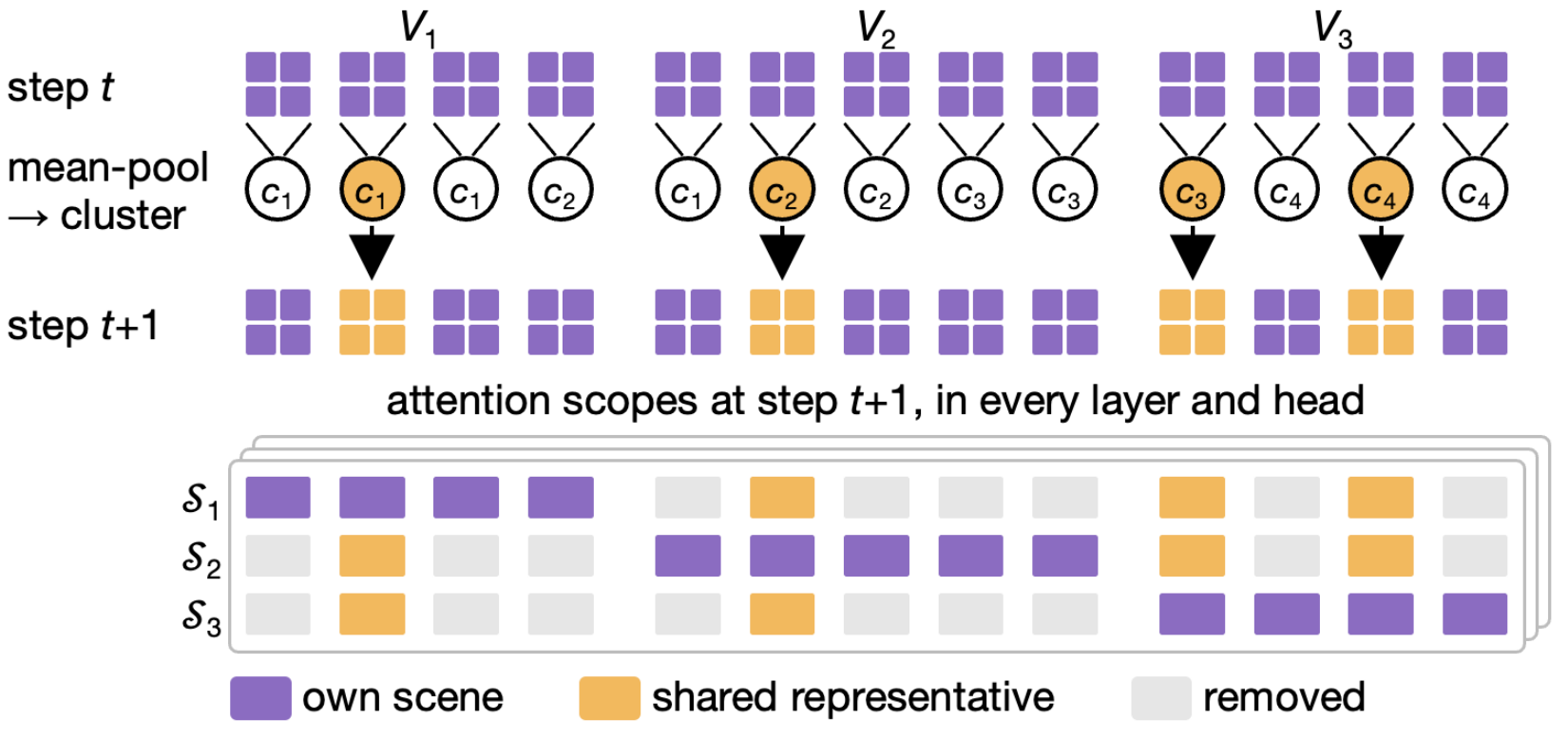}
  \vspace{-.5cm}
  \caption{Bounded cross-scene sharing. At step $t$, \name mean-pools each latent frame and $k$-means assigns it a cluster $c_k$ and the frame nearest each centroid (orange) becomes a shared representative for step $t{+}1$, used by every layer and head. The initial step uses model-specific boundary frames.}
  \label{fig:boundary-memory}
\end{figure}

As shown in Fig.~\ref{fig:boundary-memory}, \name groups these frame features into at most $M$ clusters using $k$-means and selects the original frame nearest each cluster center.
\name avoids repeated selection by reusing representative frame indices across layers and heads within a denoising step, while adapting the shared frames using the recent step's information.

\paragraph{Bounded cross-scene attention scopes.}
Let $\mathcal{R}_t$ denote the selected frame indices used at denoising step $t$, with $|\mathcal{R}_t|\leq M$. For each selected frame $f$, let $V(f)$ denote its complete set of video tokens. The shared video-token set is $
  \mathcal{M}_t
  = \bigcup_{f\in\mathcal{R}_t}V(f)$.
With $F$ video tokens per latent frame,
$|\mathcal{M}_t|\leq MF$.
Each scene $i$ accesses
$
  \mathcal{S}_i^{(t)}
  =
  V_i\cup(\mathcal{M}_t\setminus V_i)$.
Thus, each scene retains all local video tokens while accessing at most $M$ foreign frames, regardless of the number of scenes. Representatives originating from the receiving scene are excluded from its foreign context because their tokens are already available locally.

Local and shared keys participate in the same attention softmax, allowing queries to combine scene-specific information with shared context. Text and other modalities follow the architecture-specific conditioning rules described above. We use $M=4$ by default and examine the quality–latency trade-off of $M$ in \S\ref{eval:ablation}.

\paragraph{Complexity Analysis of Scene-Structured Attention.}
Dense video attention computes $L^2$ query--key pairs per head. The shared memory preserves intra-scene attention but limits each scene to at most $MF$ additional video keys. The fraction of attention work retained relative to dense attention is therefore bounded by
\begin{equation}
\frac{\sum_i \rho_iL\,|\mathcal{S}_i|}{L^2}
\leq
\sum_i\rho_i^2+\frac{MF}{L}.
\end{equation}
The first term accounts for intra-scene attention, while the second bounds cross-scene sharing. For equally sized scenes, this bound becomes $1/K+MF/L$. With scene length and $M$ fixed, both terms decrease in proportion to $1/K$, so the retained attention work grows linearly rather than quadratically with scene count.

\section{\name Runtime}
\label{sec:design}

While scene-structured attention reduces the amount of computation, realizing these savings requires efficient GPU execution. Two challenges arise. First, each scene combines dense local attention with a bounded set of foreign K/V scattered across the video sequence. Executing such irregular scopes can underutilize attention kernels. Second, scenes differ in length and attention-scope size, making equal-token GPU partitioning insufficient for balancing computation. Distributing every scene across GPUs restores balance but incurs unnecessary remote K/V accesses.

\name addresses these challenges by separating logical attention scopes from their physical execution. It packs the retained interactions into dense work (\S\ref{subsec:packing}) and employs work-balanced, scene-aware partitioning with request-level communication selection (\S\ref{subsec:balance}).

\subsection{Packing Sparse Scopes into Dense Work}
\label{subsec:packing}

\begin{figure}[t]
  \centering
  \includegraphics[width=\columnwidth]{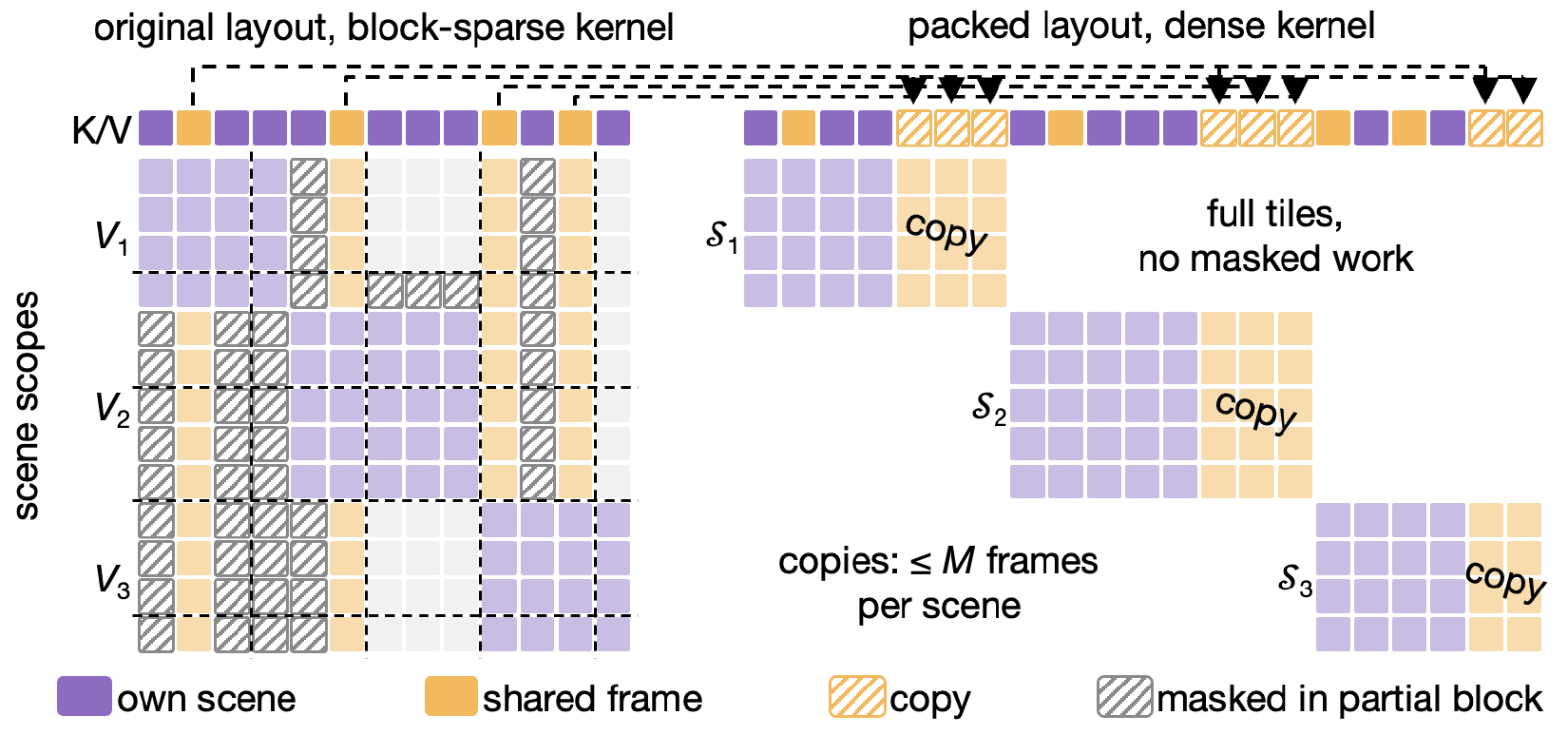}
  \caption{Packing scene scopes for dense execution. \name appends copies of shared K/V to each scene's local context, turning scattered accesses into contiguous inputs for dense variable-length attention.}
  \label{fig:packing}
\end{figure}

A straightforward way to execute scene-structured attention is to express its scopes as a block-sparse mask and invoke a sparse attention kernel~\cite{SpargeAttention,SolAttn, packinfer-arxiv26}. However, our attention scopes have a distinctive structure. Each scene retains a dense local region containing all its video tokens and accesses only a bounded number of foreign video tokens. Consequently, most retained computation occurs within dense regions, while irregularity is confined to a small shared context. This structure creates an opportunity to trade limited data replication for more efficient attention execution. Rather than preserving scattered K/V accesses through a sparse kernel, \name materializes the retained attention scopes as contiguous sequences. The additional replication is bounded by the shared-memory budget, while dense execution benefits the entire attention scope.

Specifically, \name appends copies of the required foreign K/V to each scene's local K/V (Fig.~\ref{fig:packing}). For video queries, the packed sequence contains the locally permitted keys and selected foreign video keys, excluding representatives already available locally. All video queries from the same scene reuse this prepared K/V sequence, amortizing the packing cost across their attention computation. Query classes with different key-access rules retain their architecture-specific scopes (\S\ref{subsec:scene-attention}).

Because scenes can differ in length and in the number of foreign representatives, the resulting query and K/V sequences are variable-length. \name batches the packed video-query scopes using dense variable-length attention~\cite{Flashattention3}, without padding them to a common length. The packed layout preserves the logical attention scopes and original positional information.

Table~\ref{tab:kernel-eff} compares packed execution against block-sparse execution on identical attention scopes. Including per-layer K/V copying, packed dense execution achieves $1.68$--$1.83\times$ speedup. Thus, the higher throughput of dense attention outweighs the bounded data-replication overhead. We next describe how the required K/V are distributed across GPUs.

\begin{table}[t]
  \centering
  \small
  \setlength{\tabcolsep}{4pt}
  \renewcommand{\arraystretch}{1.1}
  \caption{Speedup over block-sparse execution on identical attention scopes. Packed execution includes per-layer K/V copying.}
  \label{tab:kernel-eff}
  \begin{tabular*}{\columnwidth}{@{\extracolsep{\fill}}llccc@{}}
    \toprule
    K/V layout & Attention path & $K=3$ & $K=6$ & $K=12$ \\
    \midrule
    Original & Block-sparse & $1.00\times$ & $1.00\times$ & $1.00\times$ \\
    Packed & Dense varlen & $\mathbf{1.83}\times$ & $\mathbf{1.72}\times$ & $\mathbf{1.68}\times$ \\
    \bottomrule
  \end{tabular*}

  \vspace{3pt}
  \begin{minipage}{\columnwidth}
    \footnotesize
    \raggedright
    Block-sparse execution uses FlashInfer with FA3 for full blocks and FA2 for partial blocks. Packed execution uses FA3.
  \end{minipage}
  \vspace{-10pt}
\end{table}

\subsection{Work-Balanced Multi-GPU Execution}
\label{subsec:balance}
Scene-structured attention changes both the computational workload and the K/V dependencies of multi-GPU execution. Placing complete scenes on individual GPUs preserves local K/V but can leave computation imbalanced. Conversely, distributing every scene across GPUs balances computation but requires remote K/V exchange even for intra-scene attention. \name addresses this trade-off with a lightweight two-stage approach. It constructs work-balanced partitions that preserve complete scenes whenever possible and compares the estimated communication time of K/V gathering and Ulysses-style execution~\cite{USP}, selecting one path for the request.

\paragraph{Balancing attention work while preserving locality.}
Under dense global attention, assigning equal numbers of queries approximately balances attention computation because every query accesses the same global key sequence. This no longer holds under scene-structured attention: queries from different scenes access scopes containing different numbers of keys.

Let scene $i$ contain $n_i=\rho_iL$ video tokens, and let $c_i$ denote its estimated attention work per video query, proportional to $|\mathcal{S}_i|$. Its total video-attention workload is $w_i=n_ic_i$. Thus, equal query counts can produce unequal GPU workloads. Assigning each GPU an equal fraction of every scene's queries would balance attention work, but unnecessarily fragment scenes across GPUs, requiring remote K/V transfers even for intra-scene attention.

\name employs \emph{work-balanced contiguous partitioning}. It traverses scenes in temporal order and accumulates their estimated attention work (Alg.~\ref{alg:scene-partition}, Lines~\ref{algo:count-start}--\ref{algo:count-end}). For $P$ GPUs, it places partition boundaries at successive multiples of the target workload $\bar W=\frac{1}{P}\sum_{i=1}^{K}n_ic_i$. A scene remains intact when partition boundaries fall outside it; otherwise, only the boundaries falling within the scene split its queries (Line~\ref{algo:partition}). Because partitions are contiguous, at most $P-1$ boundaries can split scenes. This construction approximately balances computation while preserving complete scenes whenever the workload boundaries permit. It does not directly minimize communication, but avoids the systematic fragmentation caused by distributing every scene across GPUs.

Queries and their associated local K/V follow the same partition. An intact scene requires no remote K/V for intra-scene attention. For a split scene, each GPU gathers only the missing K/V needed by its assigned queries. Cross-scene representatives are exchanged when their source tokens reside on other GPUs. The resulting K/V are organized using the packed layout described in \S\ref{subsec:packing}.

\begin{algorithm}[t]
\caption{Work-Balanced Scene Partitioning}
\label{alg:scene-partition}
\begin{algorithmic}[1]
\Require Ordered scenes with length fractions $\{\rho_i\}_{i=1}^{K}$, total video-token count $L$, attention scopes $\{\mathcal{S}_i\}_{i=1}^{K}$, and $P$ GPUs
\Ensure Query ranges $\Pi_p$ assigned to each GPU $p$
\State $a_0 \gets 0$; $W_0 \gets 0$
\For{$i = 1,\ldots,K$}
    \State $c_i \gets |\mathcal{S}_i|$ \Comment{Attention work per video query}
    \label{algo:count-start}
    \State $a_i \gets a_{i-1} + \rho_i L$
    \State $W_i \gets W_{i-1} + \rho_i L c_i$ \Comment{Total attention workload}
    \label{algo:count-end}
\EndFor
\State $\bar W \gets W_K / P$ \Comment{Target work per GPU}
\State $b_0 \gets 0$; $b_P \gets L$
\For{$p = 1,\ldots,P-1$}
    \State $i \gets \min\{j : W_j \geq p\bar W\}$
    \State $r \gets \operatorname{round}\bigl((p\bar W-W_{i-1})/c_i\bigr)$
    \State $b_p \gets a_{i-1} + r$
    \label{algo:partition}
\EndFor
\For{$p = 1,\ldots,P$}
    \State $\Pi_p \gets \Call{SceneRanges}{[b_{p-1},b_p)}$
\EndFor
\State \Return $\{\Pi_p\}_{p=1}^{P}$
\end{algorithmic}
\end{algorithm}

\paragraph{Optimizing the communication path.}
The advantage of locality-aware execution depends on how scenes are distributed across GPUs. When scenes fit within individual GPUs, their intra-scene attention requires no remote K/V transfers. However, when a large scene must be partitioned to balance computation, each participating GPU must access the scene's complete K/V across all attention heads. As the scene spans more GPUs, the resulting K/V replication can incur substantial communication. Ulysses-style parallelism~\cite{USP} offers an alternative by redistributing attention tensors across heads, allowing each GPU to process the full sequence for some heads. This benefit depends on collective bandwidth, so \name compares the estimated communication time of the two paths.

We first consider video attention without shared frames. Let $D$ denote the bytes occupied by one token in one tensor across all attention heads, and let $\theta$ denote the effective bandwidth of K/V gathering relative to all-to-all. In the pure-Ulysses configuration of USP, Q, K, V, and the attention output are redistributed through all-to-all. Query-partitioned execution instead keeps queries and outputs local and gathers the missing K/V. Under the bandwidth-based cost model, gathering is preferable when
\begin{equation}
\underbrace{
2D\max_p\sum_{i:x_{p,i}>0}
\left(\rho_iL-x_{p,i}\right)
}_{V_{\mathrm{G}}}
<
\theta\,
\underbrace{
\frac{4LD(P-1)}{P^2}
}_{V_U}.
\label{eq:communication-choice}
\end{equation}
The left-hand side depends on which scenes the partition splits and how much of each remains local. The right-hand side depends on the full sequence length and GPU count. We take the work-balanced gathering path when its estimated communication is smaller and fall back to USP otherwise.

We next show how scene structure affects the choice between K/V gathering and USP. Let $S=\sum_i\rho_i^2$ and $\rho_{\max}=\max_i\rho_i$. The total attention work is proportional to $SL^2$, giving each GPU a target workload of $SL^2/P$. This budget limits how much of the longest scene can remain on one GPU, yielding a lower bound on the required remote K/V:
\begin{equation}
V_{\mathrm{G}}
\geq
2LD\left(\rho_{\max}-\frac{S}{P\rho_{\max}}\right).
\label{eq:gather-volume-bound}
\end{equation}
Combining this bound with Eq.~\ref{eq:communication-choice} gives the following necessary condition for lower communication time than USP:
\begin{equation}
\rho_{\max}-\frac{S}{P\rho_{\max}}
<
\frac{2\theta(P-1)}{P^2}.
\label{eq:gather-necessary}
\end{equation}
This condition connects the execution choice to the distribution of scene structures: \name can tolerate modest imbalance in scene structure while maintaining balanced attention workloads and preserving sufficient K/V locality to reduce communication. The derivation and communication accounting for complete attention scopes are detailed in Appendix~\ref{app:communication-bound}.

\begin{figure*}[t]
  \centering
  \includegraphics[width=\textwidth]{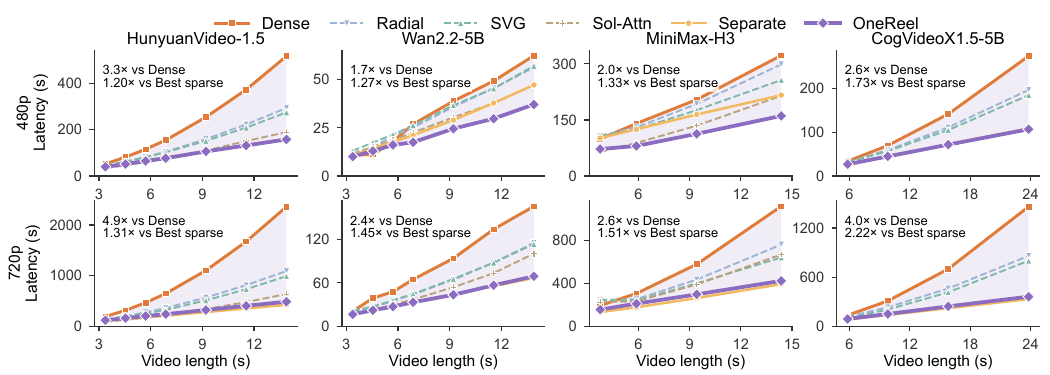}
  \vspace{-.6cm}
  \caption{End-to-end latency of multi-scene video generation against video length. Speedup numbers are computed at the longest video. Sol-Attn is omitted on CogVideoX1.5, due to kernel not support.}
  \label{fig:latency}
\end{figure*}

\section{Evaluation}
\label{sec:eval}
We evaluate \name across video generation models and multi-scene requests. Our results show that \name achieves $1.7$--$4.9\times$ end-to-end speedups over dense attention and $1.2$--$2.2\times$ over the strongest evaluated sparse-attention baseline. Meanwhile, \name also improves scene-level text alignment over dense generation.

\subsection{Evaluation Setup}

We implement a prototype of \name atop the popular video-generation engine, xDiT~\cite{xDiT}, for broad applicability. Detailed implementations are available in Appendix~\ref{app:impl}.

\paragraph{Models and Workloads.}
We evaluate \name on Wan2.2, HunyuanVideo-1.5, MiniMax H3, and CogVideoX1.5 at resolutions of 480p and 720p, using 20 denoising steps. All experiments run on 8xH200 GPUs connected by NVLink. 
We adapt multi-scene requests from HoloCine~\cite{Holocine}, Shot2Story~\cite{Shot2Story}, and MSVBench~\cite{MSVBench}, and supplement them with LLM-written scripts covering flashbacks, cross-cutting, recurring objects, and returning locations. Requests contain 3--12 scenes and specify videos up to 24\,s.

\paragraph{Baselines.}
We compare against three types of generation advances. (1)~\emph{Dense} uses each model's original full attention with FlashAttention. (2)~\emph{Separate} generates each scene independently using its own prompt and concatenates the results. (3)~\emph{Sparse attention} includes Radial Attention~\cite{RadialAttn}, Sparse VideoGen~\cite{SparseVideoGen}, and Sol-Attn~\cite{SolAttn}, using their released attention patterns, sparsity settings, and warm-up configurations. For multi-GPU execution, we additionally compare against Ulysses at matched GPU counts. Appendix~\ref{app:dbsp} further evaluates a balanced USP oracle inspired by db-SP~\cite{dbsp}

\paragraph{Efficiency metrics.}
We report end-to-end generation latency, denoising time, and per-step diffusion forward time. For multi-GPU execution, we measure data transferred per GPU and exposed communication time, together with latency scaling across GPU counts and speedup over USP. Load imbalance is measured as the maximum per-GPU attention workload divided by the average.

\paragraph{Quality Metrics.}\label{sec:metrics}
Our evaluations include popular, objective metrics and LLM as a judge for scalable evaluation. Specifically, we split videos at the planned scene boundaries. First, we use CLIP to measure the fraction of scenes whose frames best match their own prompt among all scene prompts (\emph{Text}). We additionally require cross-scene DINOv2 similarity above the 95th percentile of different-subject pairs (\emph{Text+ID}). Second, we use VBench to assess imaging quality, aesthetic quality, and dynamic degree~\cite{VBench}. Third, we use Gemini~2.5~Pro to assess scene-level action accuracy (\emph{Action}) and record blind pairwise wins/ties/losses for \name (\emph{W/T/L}).

\begin{table}[t]
  \centering
  \footnotesize
  \setlength{\tabcolsep}{3.4pt}
  \renewcommand{\arraystretch}{1.08}
   \caption{Generation quality on 40 multi-scene requests (\%, except W/T/L; metrics in \S\ref{sec:metrics}). Best and second-best scores are bold and underlined. (\emph{W/T/L}) reports \name's wins/ties/losses over the baseline.}
  \label{tab:quality}
  \begin{tabular}{@{}lcccccccc@{}}
    \toprule
    & \multicolumn{2}{c}{CLIP\,/\,DINOv2} & \multicolumn{3}{c}{VBench} & \multicolumn{2}{c}{Gemini 2.5 Pro} \\
    \cmidrule(lr){2-3}\cmidrule(lr){4-6}\cmidrule(l){7-8}
    Method & Text & Text+ID & Img. & Aes. & Dyn. & Action & W/T/L \\
    \midrule
    \multicolumn{8}{@{}l}{\textit{HunyuanVideo-1.5, 480p}} \\
    Dense    & 27.7 & 4.9  & 61.7 & 57.0 & 95.0 & 34.5 & 33/4/3 \\
    Radial   & 28.3 & 5.1  & 62.1 & 56.8 & 95.0 & 38.4 & 32/4/4 \\
    SVG      & \underline{31.2} & \underline{9.6} & 58.7 & 55.5 & \textbf{97.5} & 39.9 & 31/7/2 \\
    Sol-Attn & 30.0 & 6.9  & \textbf{67.4} & \textbf{59.4} & \textbf{97.5} & \underline{40.1} & 30/9/1 \\
    \rowcolor{gray!12}
    \name & \textbf{61.9} & \textbf{24.3} & \underline{63.2} & \underline{58.9} & \textbf{97.5} & \textbf{64.1} & NA \\
    \midrule
    \multicolumn{8}{@{}l}{\textit{Wan2.2-5B, 720p}} \\
    Dense    & \underline{25.0} & \underline{6.2} & 65.3 & 55.5 & 50.0 & 23.4 & 34/3/3 \\
    Radial   & 22.6 & 4.6  & 65.9 & 55.6 & \underline{52.5} & 27.3 & 31/7/2 \\
    SVG      & 23.5 & 3.6  & \underline{66.0} & 55.9 & \underline{52.5} & 24.6 & 37/2/1 \\
    Sol-Attn & 24.4 & 1.5  & 65.4 & \underline{56.6} & 50.0 & \underline{27.5} & 35/4/1 \\
    \rowcolor{gray!12}
    \name & \textbf{61.6} & \textbf{30.1} & \textbf{66.2} & \textbf{57.5} & \textbf{85.0} & \textbf{50.1} & NA \\
    \bottomrule
  \end{tabular}
  \vspace{-15pt}
\end{table}

\subsection{End-to-End Performance}
\label{eval:e2e}

\paragraph{\name accelerates multi-scene generation across models and resolutions.}
Figure~\ref{fig:latency} compares end-to-end latency as requests grow from 3 to 12 scenes at 480p and 720p. At twelve scenes, \name achieves $1.7$--$4.9\times$ speedups over dense attention and $1.2$--$2.2\times$ over the fastest sparse-attention baseline across the four models. Its advantage over dense attention grows with video length, increasing from $1.6\times$ to $4.9\times$ on HunyuanVideo-1.5 at 720p. 

\paragraph{\name improves scene following while preserving visual quality.}
Table~\ref{tab:quality} shows that the latency reduction is accompanied by better adherence to the requested scenes. We evaluate the compared joint-generation methods on 40 requests comprising 203 scenes using HunyuanVideo-1.5 and Wan2.2. \name achieves Text following scores of 61.6--61.9\%, compared with 23--31\% for dense and sparse-attention baselines, and also improves Text+ID. These improvements are accompanied by higher imaging and aesthetic scores than dense attention and consistent preference in blind pairwise comparisons.

\begin{figure}[t]
  \centering
  \includegraphics[width=\columnwidth]{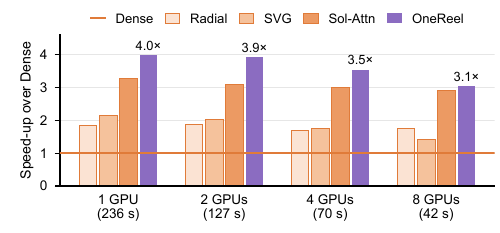}
  \caption{Speed-up of denoising over Dense attention when scaling to more GPUs.}
  \label{fig:scaling}
\end{figure}

\paragraph{\name retains its latency advantage across GPU counts.}
Figure~\ref{fig:scaling} evaluates \name's scalability on one to eight GPUs. \name is the fastest method at every GPU count, reducing denoising time from 236\,s on one GPU to 42\,s on eight. With only two GPUs, it matches dense attention on eight GPUs (127\,s versus 129\,s). Across these configurations, \name achieves $3.1$--$4.0\times$ speedups over dense attention. For this fixed request, increasing the GPU count leaves less attention work per device, making per-step overheads relatively more prominent. This limits the relative gains from attention reduction.

\begin{figure}[t]
  \centering
  \includegraphics[width=\columnwidth]{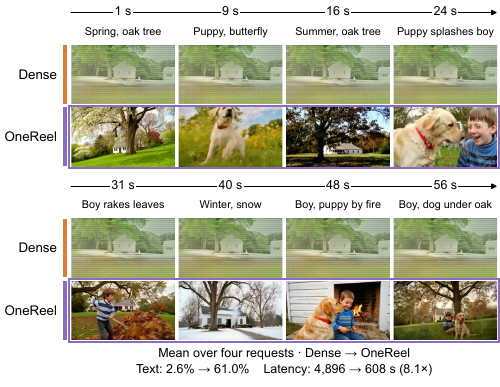}
  \caption{Minute-long generation with HunyuanVideo-1.5. \name captures the requested scene changes, while dense generation remains visually repetitive.}
  \label{fig:minute}
\end{figure}

\subsection{Long-Video Generation and Runtime Reuse}
\label{subsec:generality}
\paragraph{\name extends to minute-long multi-scene generation.}
Figure~\ref{fig:minute} illustrates generation for one-minute requests containing 26--34 scenes. Without fine-tuning, \name raises scene-level Text accuracy from 2.6\% to 61.0\%, while reducing end-to-end latency by $8.1\times$. The accompanying frames illustrate changes that dense generation fails to reproduce. Appendix~\ref{app:minute} presents the complete results and examines the contributions of scene-structured attention and RoPE.

\begin{table}[t]
  \centering
  \small
  \setlength{\tabcolsep}{4pt}
  \renewcommand{\arraystretch}{1.08}
  \caption{Runtime reuse on HoloCine with its native cross-scene attention policy.}
  \label{tab:holocine-runtime}
  \begin{tabular*}{\columnwidth}{@{\extracolsep{\fill}}lccc@{}}
    \toprule
    Execution & GPUs & Forward (s) & End-to-end (s) \\
    \midrule
    Official & 1 & 4.49 & 463.5 \\
    Ulysses & 2 & 2.45 & 261.3 \\
    \name & 2 & \textbf{2.24} & \textbf{239.8} \\
    \bottomrule
  \end{tabular*}
  \vspace{-5pt}
\end{table}

\paragraph{\name complements existing multi-scene video generation.}
We further test whether \name's runtime can accelerate native multi-scene models without changing their attention policies. We demonstrate this reuse on HoloCine~\cite{Holocine} through a thin adapter that exposes its query ranges and K/V scopes, preserving its original attention policy rather than replacing it with our bounded-memory design. As shown in Table~\ref{tab:holocine-runtime}, \name reduces forward latency by 8.6\% and end-to-end latency by 8.2\% relative to Ulysses on the same two H200 GPUs. Compared with the official single-GPU implementation, \name on two GPUs achieves a $1.93\times$ end-to-end speedup and produces an identical video after 50 denoising steps.

\subsection{Performance Breakdown}
\label{eval:breakdown}

\begin{figure}[t]
  \centering
  \includegraphics[width=\linewidth]{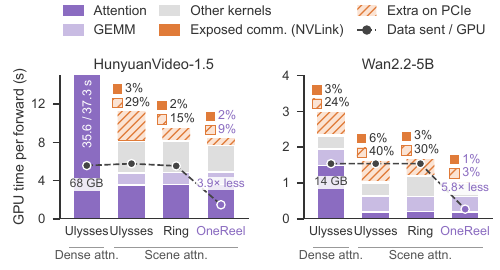}
  \caption{Per-forward GPU time and communication volume under NVLink and PCIe.}
  \label{fig:time_breakdown}
\end{figure}

\begin{figure}[t]
    \centering
    \includegraphics[width=\linewidth]{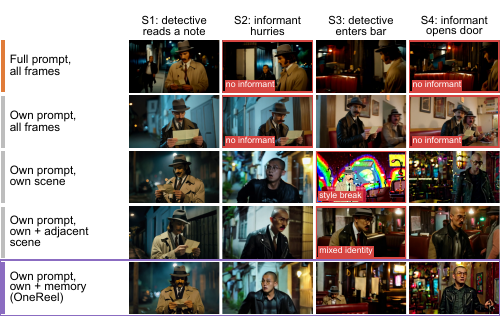}
    \caption{Visual comparison of attention scopes.}
    \label{fig:scope-examples}
    \vspace{-10pt}
\end{figure}

\paragraph{Scene-structured execution reduces attention time and K/V traffic.}
Figure~\ref{fig:time_breakdown} separates the effects of attention reduction and multi-GPU execution. Keeping Ulysses unchanged, replacing dense attention with scene-structured attention substantially reduces attention time, but leaves communication volume largely unchanged. Restricting attention scopes therefore removes computation without automatically eliminating the K/V transfers performed by conventional sequence parallelism.

On identical scene-structured scopes, \name reduces the data sent per GPU by approximately $3.9\times$ on HunyuanVideo-1.5 and $5.8\times$ on Wan2.2 relative to Ulysses. This reduction is particularly beneficial under limited bandwidth. In the PCIe setting, exposed communication accounts for 40\% of GPU time on Wan2.2 with Ulysses, compared with 3\% with \name. Under NVLink, communication occupies a smaller fraction of execution time. These results show how preserving locality benefits attention reduction.

\begin{table}[t]
  \centering
  \footnotesize
  \setlength{\tabcolsep}{4.5pt}
  \renewcommand{\arraystretch}{1.08}
  \caption{Quality under different attention scopes.}
  \label{tab:scope}
  \begin{tabularx}{\columnwidth}{@{}l>{\raggedright\arraybackslash}Xrrr@{}}
  \toprule
  Prompt & Video scope & Text & Action & Text+ID \\
  \midrule
  Full & All frames      & 27.7 & 34.5 & 4.9 \\
  Own  & All frames      & 34.9 & 31.2 & 10.3 \\
  Own  & Own scene       & 60.8 & 61.3 & 17.0 \\
  Own  & Own + neighbors & 58.0 & 55.0 & \textbf{25.4} \\
  \rowcolor{gray!12}
  Own  & Own + memory    & \textbf{61.9} & \textbf{64.1} & 24.3 \\
  \bottomrule
\end{tabularx}
\end{table}

\paragraph{Scene-local attention and bounded sharing balance scene following and consistency.}
Table~\ref{tab:scope} shows that scene-specific prompts alone are insufficient for strong scene following. With each scene attending to its own prompt, restricting video attention from all frames to its own scene raises Text from 34.9 to 60.8. The improvement therefore depends not only on assigning the correct prompt, but also on limiting the video information exchanged across scenes.

Shared memory complements this separation by providing cross-scene context. Adding it to scene-local attention raises Text+ID from 17.0 to 24.3 without reducing Text or Action. Compared with adjacent-scene sharing, memory achieves higher Text and Action at a slightly lower Text+ID score. Figure~\ref{fig:scope-examples} illustrates the role of this shared context. Local-only generation distinguishes the requested characters but introduces a style discontinuity, while memory sharing maintains a coherent appearance across the distinct scenes.

\subsection{Sensitivity Analysis and Ablation Studies}
\label{eval:ablation}

\begin{figure}[t]
    \centering
    \begin{minipage}{0.486\linewidth}
        \centering
        \includegraphics[width=\linewidth]{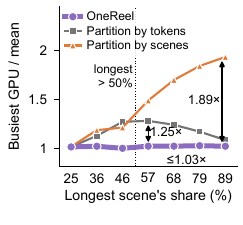}
        \caption{Attention-work imbalance across scene-length distributions.}
        \label{fig:balance}
    \end{minipage}
    \hfill
    \begin{minipage}{0.486\linewidth}
        \centering
        \includegraphics[width=\linewidth]{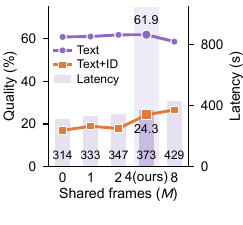}
        \caption{Effect of shared-frame budget $M$ on quality (lines) and latency (bars).}
        \label{fig:msweep}
    \end{minipage}
    \vspace{-10pt}
\end{figure}
    
\paragraph{Work-balanced partitioning tolerates unequal scene lengths.}
Figure~\ref{fig:balance} varies the longest scene's share from 25\% to 89\% on two GPUs. By accounting for per-query cost and splitting oversized scenes, Algorithm~\ref{alg:scene-partition} reduces the maximum-to-mean workload ratio by factors of up to $1.25$ and $1.89$ relative to token- and scene-based partitioning, respectively. Across all tested layouts, the maximum per-GPU workload remains within $1.03\times$ the mean.

\paragraph{Four shared frames offer a practical quality--latency trade-off.}
Figure~\ref{fig:msweep} varies the shared-frame budget from $M=0$ to $M=8$. Our default $M=4$ raises Text+ID from 17.0 to 24.3 relative to no sharing, with a 19\% increase in latency. Doubling the budget to $M=8$ provides only another 2.3 percentage points of Text+ID improvement while adding 15\% latency and reducing Text. The four-frame budget captures most of the observed Text+ID gain.

\paragraph{Impact of scene granularity and intra-scene sparsity.}
Additional experiments show that finer scene partitioning reduces latency even at a fixed video length (Appendix~\ref{app:scene-count}). Combining \name with intra-scene sparse attention provides further acceleration, particularly for longer scene-local contexts (Appendix~\ref{app:sol}).

\section{Related Work}
\label{sec:related}

\paragraph{Efficient Diffusion Inference.}
Prior work accelerates diffusion attention through low-precision kernels~\cite{SageAttention2} and spatial-temporal sparsity~\cite{SparseVideoGen,FastVideo}. Sparse VideoGen2~\cite{SparseVideoGen2} further clusters and reorders tokens to improve sparse execution. For multi-GPU inference, USP~\cite{USP} and xDiT~\cite{xDiT} provide sequence and hybrid parallelism, while PipeFusion~\cite{PipeFusion} pipelines computation across patches and model stages. DSA~\cite{DSA} combines sparse attention with distributed execution. These efforts demonstrate the importance of matching attention patterns with execution strategies. \name uses the scene structure to define attention scopes and exploits their dense local regions and bounded shared context.

\paragraph{Long and Multi-Scene Video Generation.}
StreamingT2V~\cite{StreamingT2V} and LongLive~\cite{LongLive} extend videos through autoregressive generation and context reuse, while Stable Video Infinity~\cite{StableVideoInfinity} addresses accumulated errors through error-recycling training. For multi-prompt generation, DiTCtrl~\cite{DiTCtrl} provides training-free attention control. HoloCine~\cite{Holocine} trains a multi-shot model with shot-specific text conditioning and dense intra-shot but sparse inter-shot attention. \name likewise distinguishes local scene content from shared information, but supports multiple pretrained attention architectures without retraining. 

\section{Conclusion}

\name is a training-free system for efficient multi-scene video generation. Its scene-structured attention preserves scene-specific conditioning and dense local context while bounding cross-scene sharing. The runtime packs these attention scopes for dense-kernel execution and balances attention work across GPUs while exploiting scene locality. Across four models, \name achieves $1.7$--$4.9\times$ end-to-end speedups over dense attention and $1.2$--$2.2\times$ over the fastest sparse-attention baseline. Quality evaluations show improved scene following while preserving visual quality. 

\bibliographystyle{mlsys2025}
\bibliography{main}

\clearpage
\appendix
\renewcommand{\theHequation}{appendix.\arabic{equation}}

\lstset{
    basicstyle=\ttfamily\scriptsize,
    breaklines=true,
    columns=fullflexible
}

\twocolumn[{%
  \begin{minipage}{\textwidth}
    \centering
    \includegraphics[width=\linewidth]{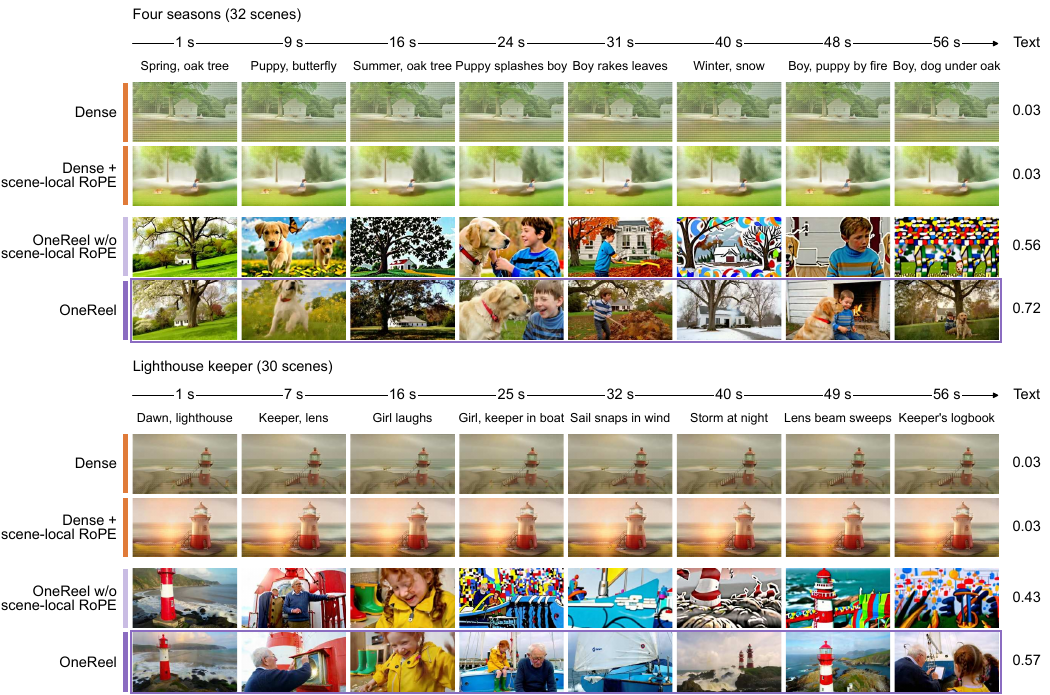}
    \captionof{figure}{Representative frames from one-minute-long requests. Scene-structured attention improves scene differentiation, while scene-local RoPE reduces late-video artifacts. Text scores are reported separately for each request.}
    \label{fig:minute-full}
  \end{minipage}
  \vspace{\baselineskip}
}]

\section{Minute-Long Multi-Scene Generation}
\label{app:minute}

We evaluate pretrained HunyuanVideo-1.5 without fine-tuning at 480p on two H200 GPUs with 20 denoising steps. Table~\ref{tab:minute-full} compares dense generation and \name, each with and without scene-local RoPE, on one-minute requests containing 1,421 frames and 26--34 scenes. Scene-local RoPE assigns Q and K temporal coordinate $\tau-a_i$, where $\tau$ is the global latent-frame index and $a_i$ is scene $i$'s starting index. Shared frames retain their source-scene coordinates regardless of the destination scene or packed position.

Dense attention exposes each query to content from all scenes, allowing unrelated content to compete with its own scene. \name restricts this context while retaining scene-specific conditioning. Figure~\ref{fig:minute-full} shows that dense generation remains visually repetitive even with scene-local RoPE, whereas \name without positional adaptation already produces distinct scenes. Across four settings, this configuration raises Text accuracy from 2.6\% to 47.2\% (Table~\ref{tab:minute-full}). These improvements in scene following do not, however, eliminate late-video degradation. Under global RoPE, a compact attention scope can still include shared keys at large temporal separations. In the illustrated examples, \name without scene-local RoPE exhibits pronounced texture and color artifacts in later frames. Scene-local RoPE adjusts these positional relationships, reduces the observed artifacts, and further raises mean Text accuracy to 61.0\% without increasing measured latency. Scene structure and positional adaptation therefore play complementary roles, improving scene following and mitigating long-horizon visual degradation, respectively.

\begin{table}[h]
  \centering
  \small
  \setlength{\tabcolsep}{3pt}
  \renewcommand{\arraystretch}{1.08}
  \caption{Effects of scene-structured attention and scene-local RoPE on minute-long generation. Scene-structured attention improves text following and reduces latency, while scene-local RoPE further improves text following without additional measured latency.}
  \label{tab:minute-full}
  \begin{tabular*}{\columnwidth}{@{\extracolsep{\fill}}l|rr@{}}
    \toprule
    Method & Text $\uparrow$ & Latency (s) $\downarrow$ \\
    \midrule
    Dense & 2.6\% & 4,896 \\
    Dense + local RoPE & 3.5\% & 4,902 \\
    \name w/o local RoPE & 47.2\% & 608 \\
    \name & \textbf{61.0\%} & \textbf{608} \\
    \bottomrule
  \end{tabular*}
\end{table}

\begin{figure}[t]
  \centering
  \includegraphics[width=\columnwidth]{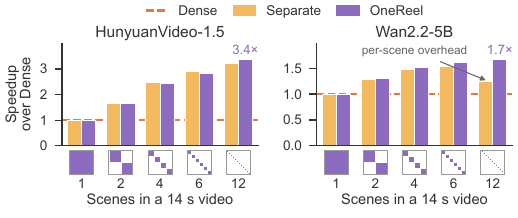}
  \caption{Speedup over dense generation as a fixed-length video (14\,s, 480p) is divided into more scenes.}
  \label{fig:scene_count}
\end{figure}

\section{Effect of Scene Numbers}
\label{app:scene-count}

We vary the number of scenes while fixing the total video length at 14\,s and the resolution at 480p. This separates the effect of scene structure from changes in video duration. As shown in Figure~\ref{fig:scene_count}, increasing the scene count from one to twelve yields a $3.4\times$ speedup on HunyuanVideo-1.5 and a $1.7\times$ speedup on Wan2.2 relative to one-scene execution. Dividing the same video into smaller scenes reduces the dense intra-scene attention workload, while cross-scene sharing remains bounded (\S\ref{subsec:sharing}).

These savings bring \name close to the latency of separate generation despite retaining cross-scene information exchange. On HunyuanVideo-1.5, its latency stays within 3\% of separate generation for 2--6 scenes. At twelve scenes, \name is faster than separate generation on both models, with a $1.34\times$ advantage on Wan2.2. Scene structure therefore provides an additional source of efficiency even when the total video length is unchanged.

\begin{figure}[t]
  \centering
  \includegraphics[width=\columnwidth]{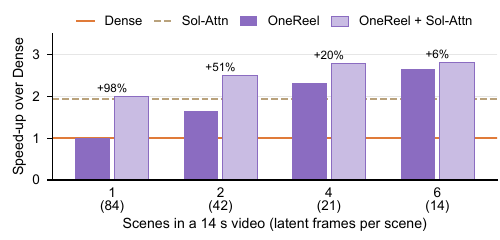}
  \caption{Combining scene-structured and intra-scene sparse attention on a fixed 14\,s video. Percentages indicate additional denoising-step speedup over \name.}
  \label{fig:sol-hybrid}
\end{figure}

\section{Combining \name with Intra-Scene Sparse Attention}
\label{app:sol}

Scene-structured attention and intra-scene sparse attention reduce computation at different levels. \name restricts cross-scene interactions but retains dense attention within each scene, leaving further opportunities for sparsification when individual scenes have long contexts. We combine the two by applying sparse attention to scene-local video keys while keeping attention to shared frames and text exact. We use Sol-Attn~\cite{SolAttn}, the best-performing sparse-attention baseline in this setting, as a representative implementation.

Figure~\ref{fig:sol-hybrid} evaluates the combination on a fixed 14\,s HunyuanVideo-1.5 video at 480p on two H200 GPUs, divided into one to six scenes. With a single long scene, \name has no cross-scene computation to remove, but intra-scene sparsification provides a $1.99\times$ denoising-step speedup over dense attention. With two to six scenes, the combination achieves $2.49$--$2.81\times$ speedups over dense attention and is $1.28$--$1.45\times$ faster than sparse attention alone.

The two techniques thus address different scene structures. For fewer, longer scenes, intra-scene sparsity provides substantial additional savings, reaching $1.51\times$ over \name with two scenes. As scenes become shorter, scene-structured attention removes more computation, and the additional gain falls to 6\% with six scenes. Combining the two extends acceleration across both long and short scene-local contexts.

\section{Comparison with Balanced Sequence Parallelism}
\label{app:dbsp}

To examine the benefit of scene locality beyond workload balancing, we compare \name against balanced Ulysses and Ring layouts, motivated by db-SP's workload-balancing approach~\cite{dbsp}. Both layouts use the same scene-structured attention scopes, frame memory, and packed FA3 kernels as \name. Ulysses partitions equally loaded heads, while Ring assigns each GPU an equal share of every scene. We call the faster layout for each request and interconnect the \emph{balanced USP oracle}, excluding planning and migration overhead.

We evaluate HunyuanVideo-1.5 at 720p on two H200 GPUs with $M=4$. We vary either request duration from 3--20\,s, averaging seven latent frames per scene, or scene count from 2--16 at a fixed duration of 10\,s (60 latent frames). Scene lengths follow evenly spaced quantiles of a log-normal distribution with $\sigma=0.5$. Each run uses three denoising steps without classifier-free guidance and reports the faster of the last two steps. We test NVLink and PCIe configurations. Equation~\ref{eq:communication-choice} selects K/V gathering throughout, with communication--computation overlap enabled for the PCIe runs.

\begin{figure}[t]
  \centering
  \includegraphics[width=\columnwidth]{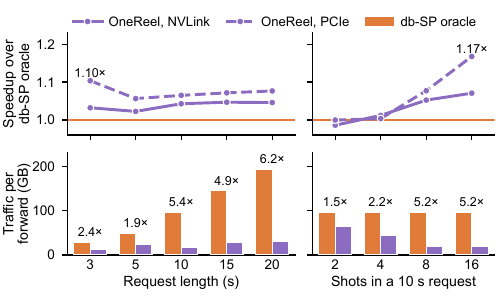}
  \caption{Comparison with the balanced USP oracle. Top: denoising-step speedup. Bottom: total communication per forward across both GPUs. Left: increasing video duration. Right: increasing scene count at a fixed duration of 10\,s.}
  \label{fig:dbsp}
\end{figure}

Figure~\ref{fig:dbsp} shows that \name reduces communication volume by $1.5$--$6.2\times$ relative to the oracle. For the fixed-length request, the oracle exchanges 96\,GB per forward regardless of scene count, whereas \name's traffic falls from 65\,GB to 18\,GB as the count increases from two to sixteen. More scenes allow the balanced partition to retain more complete scenes locally, reducing remote K/V demand without changing the total video length. The latency benefit reaches $1.07\times$ with NVLink, where Ring overlaps much of its communication with computation, and $1.17\times$ with PCIe communication. The two-scene request remains near parity ($0.99$--$1.00\times$): its longer scene contains 65\% of the frames and must be split to balance computation, limiting the locality available to \name.

\section{Communication Cost of K/V Gathering}
\label{app:communication-bound}

This appendix derives the communication comparison in Eq.~\ref{eq:communication-choice} and the scene-length condition in Eq.~\ref{eq:gather-necessary}. We first consider scene-local video attention, then account for shared frames.

\subsection{Communication Model}
\label{app:comm-setup}
\label{app:comm-volumes}

A request contains $L$ video tokens partitioned into $K$ scenes, with scene $i$ containing $\rho_i L$ tokens and $\sum_i \rho_i=1$. Attention runs on $P$ GPUs. Let $D$ denote the bytes per token in one tensor across all heads, and let $x_{p,i}$ denote the number of scene $i$'s queries assigned to GPU $p$. Their corresponding local K/V follow the same partition, so $\sum_p x_{p,i}=\rho_i L$. Without shared frames, total attention work is proportional to
\begin{equation}
  \sum_i (\rho_i L)^2 = SL^2,
  \qquad S=\sum_i\rho_i^2.
  \label{eq:app-total-work}
\end{equation}
The derivation assumes an ideal work-balanced partition with workload $SL^2/P$ per GPU. The implementation rounds partition boundaries to the supported frame granularity.

We approximate communication time by the volume received by the busiest GPU divided by effective bandwidth. Let $B$ denote the effective all-to-all bandwidth and $\theta B$ the effective K/V-gathering bandwidth, with $\theta\approx1.0$ in our measurements. This bandwidth-based model excludes collective startup latency and overlap with computation.

\paragraph{Ulysses-style execution.}
In the pure-Ulysses configuration of USP, each GPU initially holds $L/P$ tokens across all heads. All-to-all redistributes Q, K, and V into head partitions and returns the attention output to the token partition. Each of these four tensors transfers a fraction $(P-1)/P$ of its local data, giving
\begin{equation}
  V_U = 4\cdot\frac{L}{P}\cdot\frac{P-1}{P}\cdot D
      = \frac{4LD(P-1)}{P^2}.
  \label{eq:app-vu}
\end{equation}

\paragraph{Query-partitioned gathering.}
Each GPU retains all heads of its assigned queries and outputs, exchanging only missing K/V. For every scene $i$ with $x_{p,i}>0$, GPU $p$ needs $\rho_iL-x_{p,i}$ remote tokens. The busiest GPU therefore receives
\begin{equation}
  V_G = 2D\max_p\sum_{i:\,x_{p,i}>0}
  \left(\rho_iL-x_{p,i}\right).
  \label{eq:app-vg}
\end{equation}
An intact scene contributes no intra-scene communication. Under the bandwidth-based model, gathering has lower communication time when $V_G/(\theta B)<V_U/B$, or equivalently $V_G<\theta V_U$, yielding Eq.~\ref{eq:communication-choice}.

\subsection{Lower Bound and Necessary Condition}
\label{app:comm-lower-bound}
\label{app:comm-necessary}

Let $\rho_{\max}=\max_i\rho_i$. Work balance limits how many queries of the longest scene can remain on one GPU, providing a lower bound on the K/V that gathering must transfer.

\begin{lemma}
\label{lem:per-gpu-cap}
Under an ideal work-balanced partition, each GPU holds at most $SL/(P\rho_i)$ queries of scene $i$.
\end{lemma}
\begin{proof}
Each query of scene $i$ attends to $\rho_iL$ keys. The per-GPU workload budget therefore gives $x_{p,i}\rho_iL\le SL^2/P$, hence $x_{p,i}\le SL/(P\rho_i)$.
\end{proof}

\begin{proposition}
\label{prop:gather-bound}
Under an ideal work-balanced partition,
\[
  V_G \ge 2LD\left(\rho_{\max}-\frac{S}{P\rho_{\max}}\right).
\]
\end{proposition}
\begin{proof}
If the right-hand side is non-positive, the bound is immediate. Otherwise, consider a longest scene $j$. Lemma~\ref{lem:per-gpu-cap} gives $x_{p,j}\le SL/(P\rho_{\max})<\rho_{\max}L$, so the scene must span multiple GPUs. Any GPU holding its queries must receive at least $\rho_{\max}L-SL/(P\rho_{\max})$ of its tokens. Accounting for both K and V gives the bound.
\end{proof}

Combining this bound with $V_G<\theta V_U$ gives
\[
  2LD\left(\rho_{\max}-\frac{S}{P\rho_{\max}}\right)
  \le V_G
  < \theta\frac{4LD(P-1)}{P^2}.
\]
Dividing by $2LD$ yields the necessary condition in Eq.~\ref{eq:gather-necessary}:
\[
  \rho_{\max}-\frac{S}{P\rho_{\max}}
  < \frac{2\theta(P-1)}{P^2}.
\]
This condition depends only on scene lengths and the bandwidth ratio. It is necessary rather than sufficient because other split scenes can contribute additional traffic. The communication comparison therefore uses the volume of the actual partition.

\subsection{Complete Attention Scopes}
\label{app:comm-full-scope}

With shared frames, partitioning uses the full per-query attention cost $|\mathcal{S}_i|$ defined in Section~\ref{subsec:sharing}. Ulysses still redistributes the full sequence across heads, so its communication volume remains Eq.~\ref{eq:app-vu}. Gathering instead transfers the remote video K/V required by the scopes assigned to each GPU.

Let $\mathcal{L}_p$ denote the video tokens whose K/V are locally owned by GPU $p$. The complete gathered volume is
\begin{equation}
  V_G =
  2D\max_p
  \left|
    \left(\bigcup_{i:\,x_{p,i}>0}\mathcal{S}_i\right)
    \setminus\mathcal{L}_p
  \right|.
  \label{eq:app-vg-full}
\end{equation}
The union counts each remote token once per destination GPU, even when several local scenes read it. Copies subsequently created for per-scene packing do not incur additional inter-GPU traffic. Without shared frames, Eq.~\ref{eq:app-vg-full} reduces to Eq.~\ref{eq:app-vg}.

Equation~\ref{eq:app-vg-full} supplies the communication volume for the same bandwidth-based comparison. The closed-form necessary condition above characterizes scene-local attention, while complete-scope accounting includes the selected shared frames and the resulting workload partition.

\section{Implementation Details}
\label{app:impl}

\paragraph{Model integration.}
\name uses PyTorch, xDiT~\citep{xDiT}, and model-specific attention processors without modifying pretrained weights. Scene prompts are encoded independently, and HunyuanVideo-1.5 also applies its text-token refiner separately to each prompt. Model adapters expose scene ranges and modality indices.

\paragraph{Representative-frame selection.}
Frame features are spatially mean-pooled from the middle transformer block, synchronized through an all-reduce of per-frame sums and token counts, and normalized. We use spherical $k$-means with $M=4$, farthest-point initialization, and at most 20 iterations. Each cluster selects the frame nearest its centroid, breaking ties by frame index. Selected indices apply at the next denoising step, while the initial step uses model-specific boundary frames.

\paragraph{Request-level planning.}
Initial boundary scopes determine the query partition and communication path once per request. Path selection uses $\theta=1.0$, assuming equal effective bandwidths. The partition and path remain fixed throughout denoising, while representative updates change the shared-frame lists used for packing.

\paragraph{Packed attention.}
\name batches query pieces using FlashAttention-3 variable-length attention. Each layer packs K/V from its current activations. Contiguous scopes are read directly using \texttt{cu\_seqlens\_k} and \texttt{seqused\_k}, avoiding copies when the required K/V already form a contiguous range.

\paragraph{K/V and text-output exchange.}
The K/V-gathering path merges overlapping remote intervals and transfers them through batched NCCL point-to-point operations. Each remote token is received once per destination GPU. In this path, each scene's text queries are computed by the GPU holding its first video token, which sends the outputs only to GPUs that need them.

\paragraph{Communication overlap.}
In the PCIe configuration, NCCL peer-to-peer transport is disabled and transfers are staged through host memory. \name computes attention over local keys while remote K/V are transferred, then computes attention over the received keys and merges the partial outputs using log-sum-exp values. This split-attention path is disabled on NVLink.

\end{document}